# A deep reinforcement learning model based on deterministic policy gradient for collective neural crest cell migration

Yihao Zhang[1], Zhaojie Chai[1], Yubing Sun[2], George Lykotrafitis[1,3*]

[1]Department of Mechanical Engineering, University of Connecticut, Storrs, CT 06269, USA

[2]Department of Mechanical and Industrial Engineering, University of Massachusetts Amherst, Amherst, MA 01003, USA

[3]Department of Biomedical Engineering, University of Connecticut, Storrs, CT 06269, USA

*Corresponding author

**Email: george.lykotrafitis@uconn.edu**

**Abstract**

Modeling cell interactions such as co-attraction and contact-inhibition of locomotion is essential for understanding collective cell migration. Here, we propose a novel deep reinforcement learning model for collective neural crest cell migration. We apply the deep deterministic policy gradient algorithm in association with a particle dynamics simulation environment to train agents to determine the migration path. Because of the different migration mechanisms of leader and follower neural crest cells, we train two types of agents (leaders and followers) to learn the collective cell migration behavior. For a leader agent, we consider a linear combination of a global task, resulting in the shortest path to the target source, and a local task, resulting in a coordinated motion along the local chemoattractant gradient. For a follower agent, we consider only the local task. First, we show that the self-driven forces learned by the leader cell point approximately to the placode, which means that the agent is able to learn to follow the shortest path to the target. To validate our method, we compare the total time elapsed for agents to reach the placode computed using the proposed method and the time computed using an agent-based model. The distributions of the migration time intervals calculated using the two methods are shown to not differ significantly. We then study the effect of co-attraction and contact-inhibition of locomotion to the collective leader cell migration. We show that the overall leader cell migration for the case with co-attraction is slower because the co-attraction mitigates the source-driven effect. In addition, we find that the leader and follower agents learn to follow a similar migration behavior as in experimental observations. Overall, our proposed method provides useful insight on how to apply reinforcement learning techniques to simulate collective cell migration.

# 1. Introduction

Collective cell migration is the coordinated motion of a group of cells commonly observed during embryonic development and wound healing. Collective cell migration is mediated by stable cell-cell junctions and transient interactions between cells in mesenchymal tissues [1, 2]. Cells utilize various mechanisms to facilitate collective migration. Here, we are concerned with collective cell migration in mesenchymal tissues. For example, in *Xenopus,* neural crest (NC) cells migrate as a cohesive cell population driven by a distant source of stromal cell-derived factor 1 (Sdf1) secreted by placode cells [3]. This means that the directional migration of *Xenopus* NC cells is controlled by the gradient of Sdf1 concentration which leads to the shortest path to placode cells. In addition, the migrating NC cells remain a cohesive cluster via co-attraction (CoA) triggered by protein C3a secreted by themselves [4, 5]. Moreover, contact-inhibition of locomotion (CIL), which describes the repelling behavior of cells when confronting each other, also plays a critical role in maintaining the directional migration of NC cells [6, 7].

Recently, deep reinforcement learning (DRL) has been successfully implemented in areas such as robotics, control, and games [8-11]. DRL considers both policy learning and policy improvement, which has shown great potential in cell biomechanics. In a recent work, Wang *et al.* used the DRL approach on 3D sequential images to learn cell migration paths at the early stage of *C. elegans* embryogenesis [12]. In another study, Hou *et al.* speeded up collective cell migration using the DRL method [13]. They built an agent-based modeling framework to establish a simulation platform for collective cell migration using 3D time-lapse microscopy images. The goal of their modeling approach was to build a biomimetic environment to demonstrate the importance of stimuli between the leader and follower cells. However, both of them are image-based tasks from experimental observations which are usually time-consuming and computationally expensive. Also, the finite number of output of the Q-learning approach is not able to generate an infinite continuous representation for the migration direction. Overall, research employing DRL techniques on collective cell migration is very limited.

The aim of this work is to use DRL approaches to generate collective NC cell migration in mesenchymal tissues that qualitatively agrees with experimental results. We test the underlying physical principles by comparing the results obtained from the DRL agents with results obtained by numerical simulations using traditional agent-based modeling techniques. We generate data based on the principle that migrating cells follow the concentration gradient and then compare our DRL results of the velocity directions to the concentration gradient directions. We implement the actor-critic deep neural network to approximate the continuous action space and the Q-value function. The deep neural network considers cell locations and local concentration gradients of chemo-attractants or chemo-repellents as input. The actor network generates a continuous action representing the direction of a self-driven force and the critic network is used to output the action Q-value representing the corresponding cumulative rewards. The weights in the actor-critic network are updated by using a deep deterministic policy gradient (DDPG) [14]. Due to the different migration mechanisms of leader and follower NC cells, we use two types of agents to learn the migration behavior. The leader agent considers both a global target, which is the shortest path to the source of Sdf1, and a local target representing the mutual attraction induced by the C3a concentration gradient. However, the follower agent considers only the local target. Upon completion of training, the leader and follower agents produce patterns of behavior that are very similar to ground truth data.

## 2. Model and Methods

In this section, we describe the modeling methodology. First, we describe the NC cell migration simulation environment, which includes the interaction between the continuous particle dynamics motion and the concentration diffusion field. We note that the simulation environment provides the underlying physics rules of how the "world" functions but not how to play the "game". That means that the "gaming" environment itself does not provide the best policy for winning the game and a player needs to be trained in order to learn to achieve the best performance within the constraints defined by the underlying physics. In other words, the objective of the agents is to learn to reproduce similar migration behavior as those provided by the particle simulations. Here, we introduce the DDPG algorithm to train agents to find the optimal migration policy. A big advantage of DDPG is that agents are able to learn versatile policies for both leader and follower cells using low-dimensional observations with the same hyper-parameters and network structure.

### 2.1. Simulation "gaming" environment: A continuous particle dynamics model

The simulation "gaming" environment is developed by implementing a particle dynamics approach in a continuous space. For simplicity, we will use dimensionless units throughout this work since our goal is to train agents to reproduce a general migration behavior. The space is considered as a 10x10 square area that contains leader agents, follower agents, boundaries, and a target source, as shown in Fig. 1. A single agent is represented by a circle with a diameter of 0.5. The boundaries of the simulation space prevent agents to penetrate across. As agents move, the underlying physics of the simulation employs the following forces.

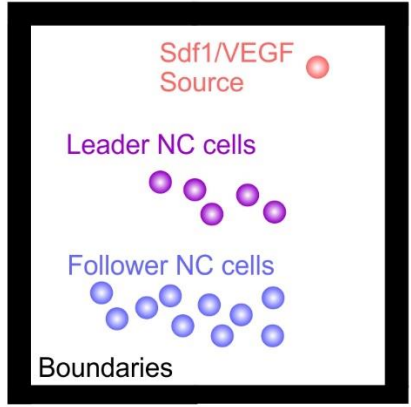

**Fig. 1**. Configuration of the particle dynamics simulation environment. The red particle represents the target source which secretes the Sdf1/VEGF chemo-attractant. The purple particles represent the leader NC cells and the blue particles represent the follower NC cells. The leader and follower cells are able to secret C3a which is responsible for CoA between NC cells. Also, CIL develops between NC cells when they collide with each other.

### 2.1.1. Self-driven force

The self-driven force represents the driving force applied to an agent in a certain direction and as a result, it changes its position accordingly. Since the direction of a self-driven force varies continuously, we consider the directional angle $\psi$ between 0 and $2\pi$. The self-driven force is given by the expression:

$$\mathbf{F}_{\text{self-driven}} = A \cdot \mathbf{f}_{\text{action}}, \tag{1}$$

where $A = 0.4|\nabla C|$ is the force magnitude proportional to the concentration gradient, and $\mathbf{f}_{\text{action}} = (\cos(\psi), \sin(\psi))$ is the unit vector representing the continuous directions of the self-driven force. We note that the parameter A in the self-driven force is calibrated to reproduce a smooth motion of agents with appropriate selection of time step as shown in the results section. In addition, the choice of force directional angle $\psi$ is compatible with the continuous output of the DDPG actor network as is discussed in detail later. The self-driven force directions are also described as the actions available to an agent.

### 2.1.2. Physical repulsive force

It is reported that NC cells experience CIL when coming into contact with each other. We simulate CIL as a repulsive force applied to agents when they approach each other closer than a specified distance. We also apply a repulsive force when agents approach the boundaries of the considered simulation space. We use an exponential expression to represent the repulsive force between agents and between agents and boundaries:

$$\mathbf{F}^{ij}_{compression} = B \cdot \exp\left(\frac{d_{ij}-r_{ij}}{E}\right) \hat{\mathbf{r}}_{ij}, \tag{2}$$

where $B = 100$, $E = 0.08$, $d_{ij} = (d_i + d_j)/2$ is the equilibrium distance between two agents i and j, $r_{ij} = |r_i - r_j|$ is the real distance between two agents, and $\hat{\mathbf{r}}_{ij} = \frac{r_i - r_j}{|r_i - r_j|}$ is the unit vector pointing from j to i. When the distance between agents and between agents and boundaries is smaller than the equilibrium distance ($r_{ij} < d_{ij}$), compression generates a repulsive force. We note that this soft exponential compression force allows small penetration at the location of the collision and it is large enough, as $r_{ij}$ becomes smaller, to prevent agents from overlapping. In the case of boundaries, a similar exponential repulsive force also applies to an agent but in this case the direction of the repulsive force is perpendicular to the contact surface pointing towards the agent and away from the boundary.

### 2.1.3. Viscous damping force

If only a constant self-driven force is applied to an agent, then the agent may accelerate continuously for a significant time interval and reach an unrealistically high speed before it collides with another agent or an obstacle violently. In addition, the cell-substrate adhesion in the rear of the cell mitigates the migration velocity. Because of this, we hinder unrealistically large velocities by introducing the viscous damping force:

$$\mathbf{F}_{viscous} = -\frac{m}{\tau}\mathbf{v}, \tag{3}$$

where m=1 is the mass, $\mathbf{v}$ is the velocity of an agent, and $\tau = 0.5$ is the relaxation time corresponding to the viscous force.

In our simulations, the target source is stationary (Fig. 1). Also, the parameters for each agent can vary but we choose, without loss of generality, identical values for all agents in order to reduce the number of variables. The motion of an agent follows the Newton equation:

$$m\frac{d^2\mathbf{r}}{dt^2} = \sum \mathbf{F}_i, \tag{4}$$

where $\mathbf{r}$ is the position and $\mathbf{F}_i$ is the total force acting on an agent $i$. For each time step, the position ($\mathbf{r}$), velocity ($\mathbf{v}$) and acceleration ($\mathbf{a}$) are updated using the leapfrog algorithm:

$$\mathbf{v}\left(t + \frac{\Delta t}{2}\right) = \mathbf{v}(t) + \mathbf{a}(t)\frac{\Delta t}{2}, \tag{5}$$

$$\mathbf{r}(t + \Delta t) = \mathbf{r}(t) + \mathbf{v}\left(t + \frac{\Delta t}{2}\right)\Delta t, \tag{6}$$

$$\mathbf{v}(t + \Delta t) = \mathbf{v}\left(t + \frac{\Delta t}{2}\right) + \mathbf{a}(t + \Delta t)\frac{\Delta t}{2}, \tag{7}$$

where t is the current time and $\Delta t = 0.1$ is the time step.

## 2.2. Diffusion model

The diffusion model describes the concentration field in the simulation environment and it is coupled with the particle dynamics motion. As the agents move, they act as multiple moving sources of chemo-attractants affecting the distribution of the concentration field, which in turn determines the motion of the agents. Then, the interaction between particle motion and concentration space requires that the simulation environment solves the diffusion equation iteratively as a result of new locations of the agents.

We assume that diffusion occurs in the same 2D space where the particles are moving. To determine the concentration distribution in the simulation environment, we solved the parabolic equation

$$\frac{\partial C}{\partial t} = D \left( \frac{\partial^2 C}{\partial x^2} + \frac{\partial^2 C}{\partial y^2} \right), \tag{8}$$

where $D$ is the diffusion coefficient and $C$ is the concentration. The boundary conditions are defined as impenetrable barriers with $\nabla C \cdot \hat{\mathbf{n}} = 0$. The initial conditions are defined at $C_0 = 1$ for all source agents $C_0 = 0$ for all other places. During the simulation, the concentrations at the locations of the agents remain constant meaning that the agents act as sources of the secreted substances.

To solve the diffusion equation, we employed a forward-time central-spaced scheme

$$C_{i,j}^{t+1} = C_{i,j}^t + \Delta t \cdot D \left( \frac{C_{i+1,j}^t - 2C_{i,j}^t + C_{i-1,j}^t}{\Delta x^2} + \frac{C_{i,j+1}^t - 2C_{i,j}^t + C_{i,j-1}^t}{\Delta y^2} \right), \tag{9}$$

where $C_{i,j}^t$ is the concentration at the node with indices $(i, j)$, the superscript index t determines the time step, and $\Delta x = \Delta y = 0.1$ are the distances between the nodes in the x and y directions, respectively. The diffusion time step $\Delta t$ was determined by as $\Delta t \ll \Delta \mathrm{x}^2/4D$ to ensure the stability of the algorithm. Here, the diffusion coefficient is 1 and $\Delta t \ll 0.0025$. We choose $\Delta t = 0.001$ for the calculation of the diffusion equation.

### 2.3. Deterministic policy gradient algorithm for deep reinforcement learning

The fundamental elements in reinforcement learning (RL) are: state (s), action (a), reward ($R$), and policy ($\pi$). By state we mean the information that represents the environment. For example, the positions of leaders and followers and concentration gradient direction. The state of environment is used to generate an observation and a reward that are transmitted to the agent. The state of an agent is its internal representation of the environment. An agent must be able to sense the state of the environment to some extent. For instance, in the case of a fully observable environment, the agent directly observes the environment and it can access all the information of the environment. In most cases, however, an agent only partially observes the environment. Specifically, in this work we introduce two types of agents, leaders and followers. The leaders move along the concentration gradient of Sdf1, secreted by placode cells. We will first train a single leader agent learning to find the shortest path to the placode source. Then we consider the effect of C3a secreted by other leader NC cells and the overall target of the leader agent is to learn a cohesive clustered migration behavior. The followers follow the concentration gradient of C3a secreted by the leaders and other followers forming a collective NC cell migration.

In this work, the state of all types of agent is represented by the position $(x, y)$ of an agent, and the concentration gradient direction $(c_x, c_y)$ at $(x, y)$ in the 2D simulation environment. Interaction between the agent and the environment is described as a sequence of discrete time steps. At time t, the agent selects an action ($a_t$) based on the state ($s_t$) and transitions to the next state ($s_{t+1}$) where it receives a reward ($R_{t+1}$). The sequence ($s_t, a_t, R_{t+1}, s_{t+1}$) continues until the end of each episode. The goal in RL control is to identify the optimal policy for an agent at various states leading to a maximum cumulative reward. The resulting Bellman optimality equation for the optimal action value function $Q_*(s, a)$ is given by the solution of the equation:

$$Q_*(s, a) = \sum_{s', R} p(s', R|s, a)[R + \gamma \times \max_{a'} Q_*(s', a')] \geq Q_\pi(s, a), \forall \pi,$$

where $p(s', R|s, a)$ is the probability that the agent moves to a state $s'$ from a state $s$ after receiving a reward R and taking an action a. $0 \leq \gamma \leq 1$ is the discounting ratio representing the importance of future rewards.

A popular methods to obtain $Q_*(s, a)$ is the model-free Q-learning algorithm defined by Watkins et al. [15],

$$Q(s, a) = Q(s, a) + \alpha[R + \gamma \times \max_{a'} Q(s', a') - Q(s, a)],$$

where $\alpha$ is a parameter determining the updating speed. In the Q-learning algorithm, the learned Q-value directly approximates the optimal action value function by a one-step increment from the maximum $Q(s', a')$ in the next state. The converged $Q(s, a)$ is the optimal action value function $Q_*(s, a)$.

The Q-learning algorithm significantly simplifies the learning process and it is a widely implemented technique in RL. In particular, in the case of high dimensional input state, the action value function is approximated by a deep neural network (DNN) [9, 16]. The parametrized action value function $Q(s, a; w)$ takes the state (s) as input and outputs action values through a DNN with trainable parameters w. In recent years, Mnih et al. introduced two major techniques: memory relay buffer and target network update to overcome the instability of direct implementation of deep neural networks into RL [9]. They showed that agents trained using the Q-learning algorithm with deep neural networks (DQN) are able to beat human-level scores in several Atari games. Although, DQN has emerged as a popular DRL algorithm and it has achieved great success in games, robotics and controls, implementation of continuous action space into the DQN is not possible. The main reason is that the Q-learning algorithm finds the greedy action with the maximum cumulative rewards for each step, but this optimization is too slow to be practical in a continuous action space. Here, we employed the DDPG algorithm which is able to handle continuous action spaces efficiently [14].

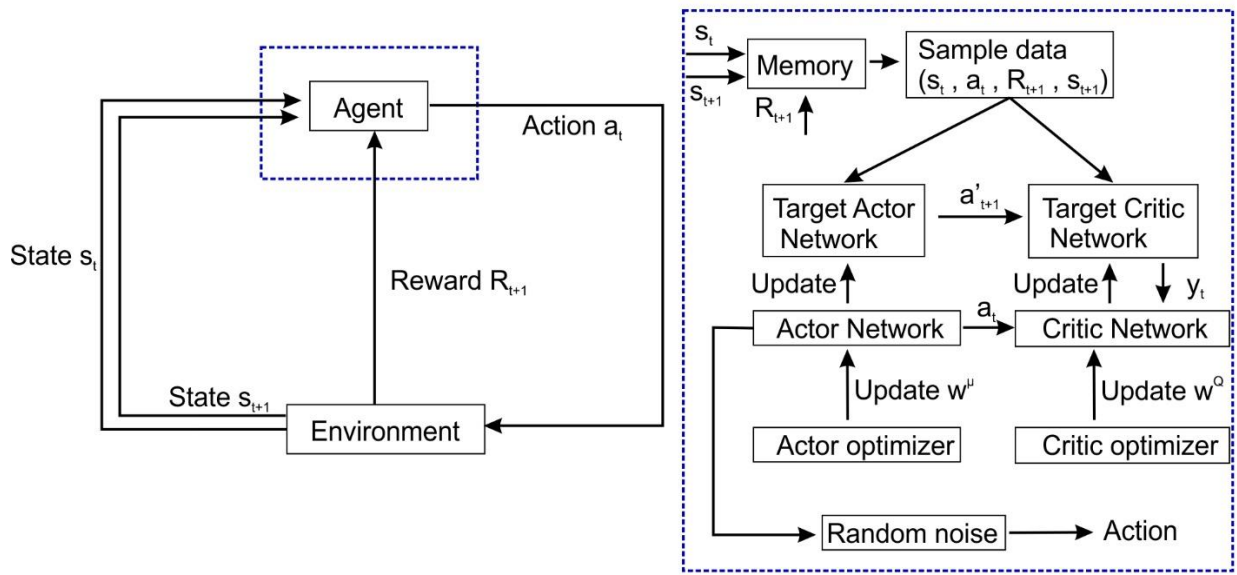

**Fig. 2**. DDPG Training Diagram. The input state $s_t$ is used to generate a deterministic continuous action $a_t$ through the actor network. The output action is concatenated with the input state to find the expected action value $Q(s, a)$ through the critic network. The weights of critic network are first updated by the reward and the action value of the next state. Then the weights in the actor network are updated based on the corresponding action value.

The DDPG is an actor-critic approach where the actor network is used to approximate the action $a(s, w^\mu)$ and the critic network representing the action value function $Q(s, a)$. It generates a specific action through the current state and a parameterized actor function deterministically. It has been proven that the deterministic policy gradient is the gradient of action performance [17]. The actor network is then updated by applying the chain rule as:

$\nabla_{w^\mu} J = E[\nabla_{w^\mu} Q(s, a, w^Q)] = E[\nabla_a Q(s, a, w^Q) \nabla_{w^\mu} a(s, w^\mu)]$, where $w^\mu$ and $w^Q$ are the parameters in the actor network and critic network respectively, as shown in Fig. 2. The critic $Q(s, a)$ is updated using the Bellman equation. Fig 3. shows the details of training and updating procedure in the DDPG algorithm.

---

**DDPG Learning Algorithm**

(a) Initialize the environment: simulation size, source and NC cells locations.
(b) Initialize the algorithm's parameters such as maximum training episodes, maximum steps per episode, and learning rate.
(c) Initialize the networks for the agent: actor network $a(s, w^\mu)$ and critic network $Q(s, a, w^Q)$ with $w^\mu$ and $w^Q$ randomly initialized.
(d) Initialize target actor and critic network $a'$ and $Q'$ with weights $w^{\mu'} = w^\mu$ and $w^{Q'} = w^Q$.
(e) **FOR** each training episode:
    Initialize a random process N for action exploration and an initial state $s_0$.
    **FOR** each step:
        Choose an action ($a_t$) at the current state ($s_t$) from the current policy and exploration noise.
        Obtain the reward ($R_{t+1}$) and the next state ($s_{t+1}$).
        Store the state transition sequence ($s_t, a_t, R_{t+1}, s_{t+1}$) into the memory.
        When the memory reaches the maximum length:
            Select a mini batch of samples from replay memory randomly.
            Set $y_t = R_{t+1} + \gamma Q'(s_{t+1}, a'(s_{t+1}, w^{\mu'}), w^{Q'})$
            Update the critic network by minimizing the loss:
            $L = E[(y_t - Q(s_t, a(s_t, w^\mu), w^Q))^2]$
            Update the actor network using the sampled policy gradient:
            $\nabla_{w^\mu} J = E[\nabla_{w^\mu} Q(s, a, w^Q)] = E[\nabla_a Q(s, a, w^Q) \nabla_{w^\mu} a(s, w^\mu)]$.
            Update the target network from the training network.
            $W_{target} = W_{target} + \varphi(W_{train} - W_{target})$
    **END** of episode steps
**END** of training episode

---

**Fig. 3. DDPG learning algorithm.**

There are several key points in a training session that we would like to mention. In this work, we consider the migration of two types of NC cells: leaders and followers. A leader is attracted by the chemo-attractant target Sdf1 and it heads to the source location following the shortest distance path. A leader, during its migration, secretes the chemo-attractant C3a, whose concentration field induces a driving force for the leaders and followers. In addition, a follower secretes C3a that acts as chemo-attractant for the other followers and the leaders. Combination of the effects of Sdf1 and C3a results in formation of groups of leader cells and follower cells that remain close to each other during migration. Due to different motion mechanisms involved in the migration of leaders and followers, we consider two types of learning agents, the leaders and the followers. As a leader moves, a negative reward -0.1 is given at each time step unless it reaches the target, when a 0 end reward is given. We choose the discounting factor $\gamma = 1.0$ to ensure a strong influence of future rewards. As a result, the longer the leader takes to reach the target, the lower the reward is. Thus, the choice of rewards are compatible with the goal for fast target reaching since from a RL perspective the maximum cumulative reward is equivalent to the minimum number of total time steps. On the other hand, we do not provide a step reward for the followers but a local reward $R_{local} = 1.0 \times \exp(-\beta^2) - 1$, where $\beta$ is the angle between the action direction and the local C3a concentration gradient. It is clear that smaller value of $\beta$ gets more reward. The discounting factor $\gamma = 0$ is used for the follower agents since the reward is only determined by the local angular difference. Note that the overall reward for the leader NC cells is a combination of the step and local reward. Details will be given in the results section.

In RL, an agent finds the optimal policy by learning from experience. Then it is important to have sufficient exploration, which is usually a major challenge in continuous action space. Here, we choose an Ornstein-Uhlenbeck process $dN_t = -\theta N_{t-1} + \delta dW_t$, $N_t = N_{t-1} + dN_t$, where $\theta = 0.15$, $\delta = 0.2$, and $dW_t$ is a standard normal distribution. It is used to generate a random noise to the current policy with inertia [18]. The corresponding exploration policy is given by $a'(s) = a(s, w^\mu) + N$, where N is the noise process to the actor policy. We note that an important technique for updating the trainable weights is to use a train-target network configuration. The parameters $w_{target}$ are updated from $w_{train}$ by a factor $\varphi$:

$$w_{target} = w_{target} + \varphi(w_{train} - w_{target}). \tag{14}$$

$\varphi$ ranges from 0 to 1 signifying the rate of network update. When $\varphi = 0$, the target network does not update. When $\varphi = 1$, the target network updates directly to the newly trained network. We choose $\varphi = 0.1$ in this work in order to alleviate the problem of instability while maintaining a relatively fast updating rate [8, 9]. Another important technique in the training session is the experience replay used for model-based planning. At each time step, we store the current state, action, reward, and next state $(s_t, a_t, R_{t+1}, s_{t+1})$ into a replay memory for model learning. During a training step, a random mini batch of these stored tuples is chosen to update the weights of the training network. This method avoids a strong correlation within a sequence of observations and reduces the variance between updates. We employ the Adam optimizer [19] with a learning rate of $1 \times 10^{-4}$ to minimize the loss function. The weights in the network are updated via TensorFlow 1.12.0 [20] and its reinforcement learning library TRFL 1.0.0. For the visualization of the evacuation, we use Ovito 2.9.0 [21].

## 3. Results and Discussion

### 3.1. Directional migration of NC cells by Sdf1

We first study the migration of leader NC cells induced by Sdf1. In our simulations, we assume that a placode cell is a fixed target source and it expresses a stationary concentration field of Sdf1. As a consequence, the NC cells are attracted by the placode in a direction that results in the shortest distance. Then, the aim of the leader NC migration agent is to learn the direction resulting in the shortest distance to placodes using its location (x, y) and the local Sdf1 concentration gradient (cx, cy). The leader NC agent considers these four variables as input and outputs a value within a range of [-1, 1) through the hyperbolic tangent activation function, which is further mapped to [0,2π) showing the direction of self-driven force.

The actor network has 3 hidden layers with 64, 128, and 64 neurons and an output layer with 1 neuron. The critic network imports position, concentration gradient direction, and action value from the output layer of the actor network as input and finds the Q-value through another 3 hidden layers with 64, 128, and 64 neurons and a final output layer with 1 neuron. The weights in the hidden layers are initialized with HE normalization [22] and the rectified linear unit (ReLU) function is used as the activation function. The position of an agent is rescaled to (-0.5, 0.5) and the Sdf1 concentration gradient is normalized to accelerate the learning speed.

As we mentioned in the previous section, we train the network to identify the optimum direction of the self-driven force $\mathbf{F}_{self-driven}$ depending on the agent's position and the concentration gradient. In the early stage of training episodes, due to the random noise of the Ornstein-Uhlenbeck process, an agent will be moving randomly until it reaches the placode. We set the maximum time steps for each episode to 10,000 allowing a sufficient exploration experience for the agent. The Q-values of states far away from the end state are learned from the corresponding next states in a recursive manner due to the nature of the one-step updating algorithm for the critic network and actor network.

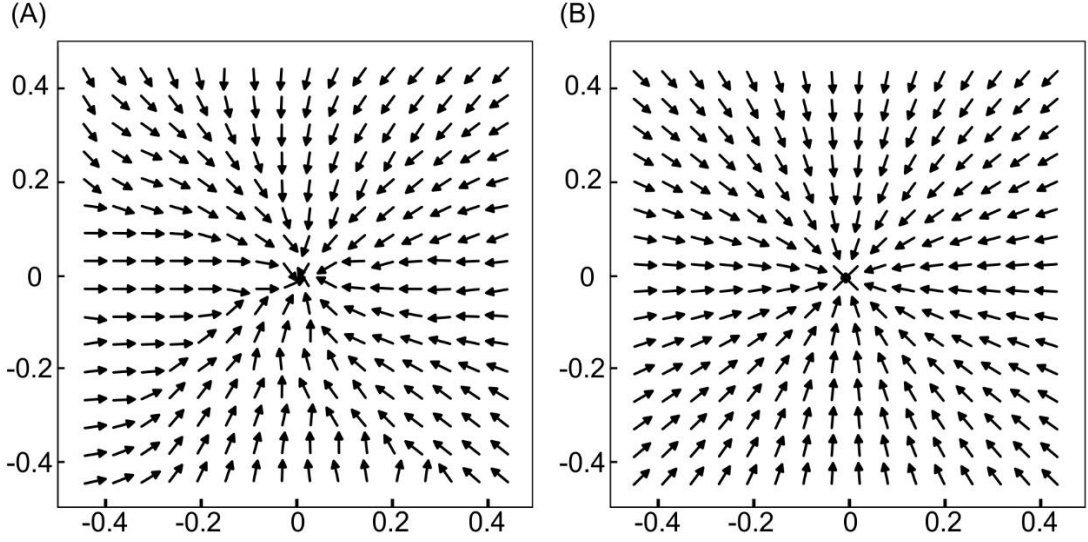

**Fig. 4**. Distribution of self-driven force directions from the DDPG and the numerical solution of the diffusion equation. (A). The learnt self-driven force directions using the DDPG algorithm. (B) Theoretical solutions of the desired self-driven force directions. The source is located in the center and the DDPG algorithm generates a very similar distribution pattern as the theoretical solution.

When the training is completed, the optimal policy is to choose the direction of self-driven force which has the maximum Q-value for a certain state $(x, y, c_x, c_y)$. Fig. 4A, shows the distribution of self-driven force directions as a result of the decision made by the agent according to the positions and concentration gradient directions. We find that the self-driven forces point approximately to the placode, which means that the agent is able to learn to approximately follow the shortest path to the target. To validate the results of the DDPG approach quantitatively, we compare the learned directions of the $\mathbf{F}_{\text{self-driven}}$ to the desired directions that directly point to the source as shown in Fig 4B. The desired directions were generated by numerically solving the diffusion equation and then computing the gradient of the obtained concentration. The efficiency of NC cell migration is evaluated by the number of time steps needed for an agent to reach the target placode cell starting from a random initial position, as shown in Fig. S1. We compare the results of the DDPG and numerical solution of the diffusion equation by plotting the histograms of the time steps for 10,000 episodes. The average number of time steps for the theoretical solution and agents learned from DDPG are 160.03 and 158.87 respectively. The corresponding variance are 6.26 and 5.83. By applying the non-parametric Mann-Whitney U test on the two data sets, we found that the p-value is $0.148>0.05$ meaning that there is no significant difference between the results produced by the desired solution and by the proposed DDPG model. We conclude that an NC cell agent is able to learn from the DDPG approach instead of following a simple rule to obtain an optimal NC cell migration induced by Sdf1.

### 3.2. Collective leader NC cells migration with CIL and CoA

We have shown that a single NC agent is able to learn the direction of migration which is controlled by the concentration gradient of Sdf1. Next, we study the collective migration of leader NC cells as a group while they obey CIL and respond to the chemoattractant C3a-mediated CoA. Two cases are considered here: 1. CIL without CoA; 2. With both CIL and CoA.

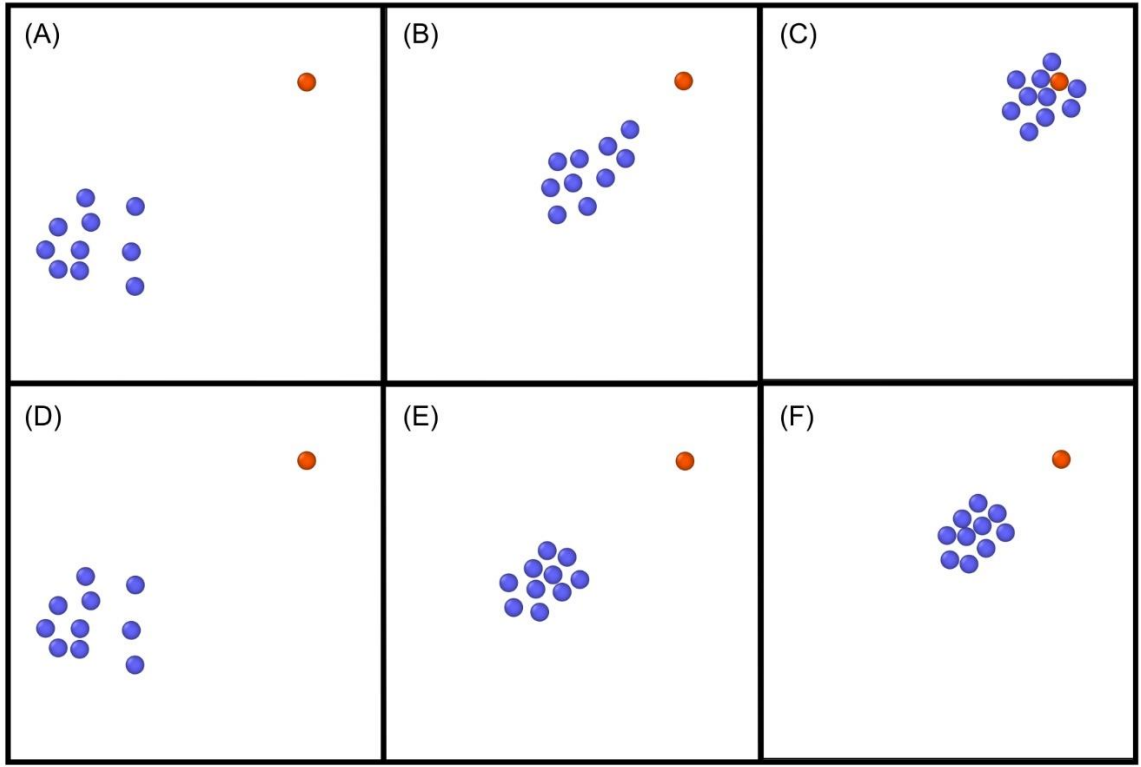

**Fig. 5**. Migration of leader NC cells to the placode with CIL and without CoA (A) to (C), with CoA (D) to (F).

In the first case, NC cells do not secrete C3a while CIL provides repulsive force when NC cells come in contact with each other. Because the Sdf1 concentration is the only source attracting the NC cells, we can employ the network trained above directly to multiple NC agents. As shown in Fig. 5A to C, the NC cells migrate towards the source along the shortest path. However, when they come closer, they push each NC cell away. The results show a collective NC cell migration in the existence of CIL but not CoA.

Next, we study the case where cell experience both CIL and CoA. Due to the effect of local C3a concentration, NC cell migration will be directed not only by the gradient of Sdf1, but it will also be affected by the mutual attraction force induced by the C3a. In order to consider the effect of both the Sdf1 and C3a, we use the idea of "motion imitation", meaning that the local C3a concentration effect controls the "motion" of the NC agent, whereas the Sdf1 represents a global task. The general learning objective is to train an agent to imitate the local "motion" induced by C3a attraction but also to reach the target source as fast as possible. Then the overall reward has two parts: a local reward $R_{local} = 1.0 \times \exp(-\beta^2) - 1$, which defines how well the agent imitates the C3a effect, and a step reward $R_{step} = -0.1$, which controls how fast the agent will reach the source. We consider the total reward $R_{total}$ at each step as $R_{total} = \omega_l R_{local} + \omega_s R_{step}$, where $\omega_l$ and $\omega_s = 1 - \omega_l$ are the weights. The weights determine the counterbalance between local C3a concentration gradients and the global Sdf1 source effect. When $\omega_l = 0$ only the global placode source is considered, which is the same situation shown in the previous case. The agent learns to find the shortest path to the placode source. On the other hand, when $\omega_l = 1$, agents consider the effect of local C3a concentration gradients only, resulting to accumulation of the NC cells into groups instead of migrating to a specific target. In the present case, we choose $\omega_l = 0.3$ in order to emphasize a strong global migration motivation while maintaining a relatively stable grouping effect induced by the local C3a concentration gradients. After 100,000 training episodes, we compare the trajectories of the agents in both cases in the same initial configuration, as shown in Fig. 5, and Video S1 and S2 in the supporting material. Due to the existence of C3a, the NC cells accumulate and stay close to each other instead of competing to reach the target source as soon as possible. As a consequence, the clustered NC cells migrate towards the placode as a group. We note that the overall migration is slower than in the case without C3a acting as chemo-attractant between NC cells because the CoA mitigates the source-driven effect.

### 3.3. Collective leaders and followers NC cells migration

It has been reported that a small number of NC cells at the migration front respond to the gradient of the Sdf1 [23, 24]. These leader cells direct the collective motion of NC cells when the follower cells are directly in contact with them. A possible explanation is that the gradient of Sdf1 is consumed by the leader NC cells, and as a result, the migration of follower NC cells are mainly directed by the C3a secreted by both the leaders and the followers [25]. Based on this observation, we consider that the migration of leader NC cells is driven by both the global Sdf1 secreted from the placodes and the C3a secreted by other leader cells. For the followers NC cells we consider only the effect of the C3a secreted by other NC cells.

For the migration of leader agents, the collective motion is controlled by both the Sdf1 and C3a. Because of this, we directly transfer the actor-critic networks from previous and apply the same weights for the total reward $R_{total}$, where $\omega^l = 0.3$. Following a similar training approach, the aim of the leader NC cells is to migrate to a target source as trailblazers directing the collective motion of all NC cells. The follower cells, however, comply with a different migration mechanism forming clusters and then follow the guidance of leader cells under the effect of C3a concentration gradients. Since there is no explicit global task, we choose $\omega^l = 1.0$ meaning that the goal of followers is to respond to the local concentration effect. The overall target of the leaders and followers is to learn to migrate in a collective manner when the migration direction is determined by the global Sdf1 and clustering is controlled by the C3a concentration.

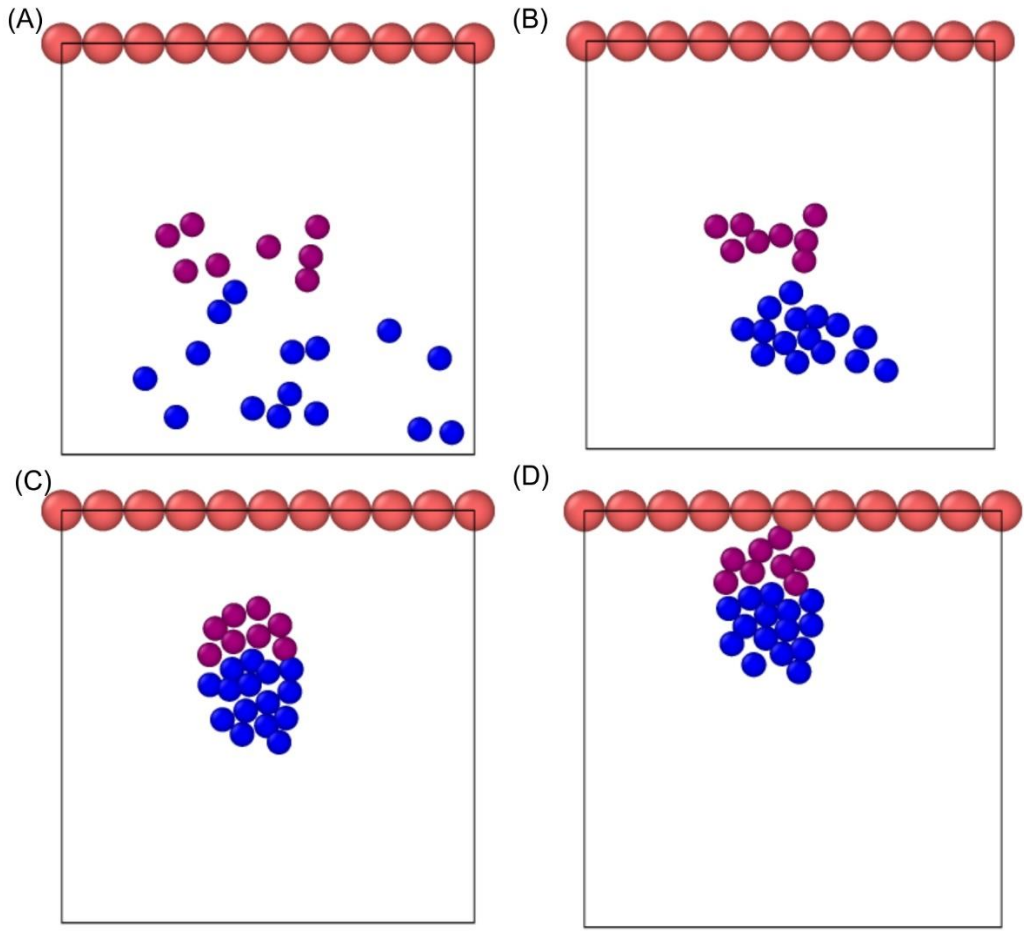

**Fig. 6**. Collective NC cells migration learned from DDPG leader and follower agent. The red particles on the top boundary are the target Sdf1 source. The purple and blue particles are the leaders and follower NC cells, respectively. (A). The initial distribution of NC cells. (B). Both leader and follower cells are grouped during the early stage of migration. (C). The leader NC cells act as the trailblazers driving the collective migration. (D). The NC cells reach the target Sdf1 source representing the final stage of NC cell migration.

The leader and follower are trained separately with the same (100,000) training episodes. The corresponding results of migration trajectories from simulation and DDPG agents are shown in Fig. 6 and Video S3. We find that both the leader and follower NC cells first group together as shown in Fig. 6B. Then the leader NC cell group act as a trailblazer driving the follower NC cell group to the source target, as shown in Fig 6 C and D. Video S4 to S7 show that the initial distribution of NC cells does not affect the overall collective migration behavior.

## 4. Conclusions

In this paper, we develop a novel DRL learning approach for collective NC cells migration in a particle dynamics environment. We implement the actor-critic DNN to approximate the continuous action space and the Q-value function. The DNN considers the location and local concentration gradient as input and outputs a continuous value of action representing the direction of self-driven force as well as a Q-value representing the corresponding cumulative rewards. The weights in the actor-critic network are updated by the DDPG learning approach. The leader agent considers a linear combination of a global task, resulting in a migration along the shortest path to the target source of Sdf1, and a local task resulting in a motion along the local C3a concentration gradient. A follower agent, however, considers only the local task. When the training is completed, the leaders and followers learn to follow collective migration trajectories with directions similar to the directions of the concentration gradient obtained by the numerical solution of the diffusion equation. In addition, they learn to form separate leader and follower groups based on CIL and C3a induced co-attraction. The proposed DRL model successfully reproduces the collective NC cells migration behavior and it provides useful insight in applying reinforcement learning techniques to predict biophysical behaviors.

**Acknowledgements**

This work used the computing resource on the San Diego Supercomputer Center (SDSC) supported by the Extreme Science and Engineering Discovery Environment (XSEDE).

**References**

1. Costa, M.L., et al., *Distinct interactions between epithelial and mesenchymal cells control cell morphology and collective migration during sponge epithelial to mesenchymal transition.* J Morphol, 2020. **281**(2): p. 183-195.
2. Theveneau, E. and R. Mayor, *Collective cell migration of epithelial and mesenchymal cells.* Cell Mol Life Sci, 2013. **70**(19): p. 3481-92.
3. Theveneau, E., et al., *Chase-and-run between adjacent cell populations promotes directional collective migration.* Nature cell biology, 2013. **15**(7): p. 763-772.
4. Carmona-Fontaine, C., et al., *Complement fragment C3a controls mutual cell attraction during collective cell migration.* Dev Cell, 2011. **21**(6): p. 1026-37.
5. Szabo, A. and R. Mayor, *Modelling collective cell migration of neural crest.* Curr Opin Cell Biol, 2016. **42**: p. 22-28.
6. Carmona-Fontaine, C., et al., *Contact inhibition of locomotion in vivo controls neural crest directional migration.* Nature, 2008. **456**(7224): p. 957-61.
7. Theveneau, E., et al., *Collective chemotaxis requires contact-dependent cell polarity.* Dev Cell, 2010. **19**(1): p. 39-53.
8. Mnih, V., et al., *Asynchronous Methods for Deep Reinforcement Learning.* 2016.
9. Mnih, V., et al., *Human-level control through deep reinforcement learning.* Nature, 2015. **518**: p. 529.
10. Silver, D., et al., *Mastering the game of Go with deep neural networks and tree search.* Nature, 2016. **529**(7587): p. 484-9.
11. Silver, D., et al., *Mastering the game of Go without human knowledge.* Nature, 2017. **550**(7676): p. 354-359.
12. Wang, Z., et al., *Deep reinforcement learning of cell movement in the early stage of C.elegans embryogenesis.* Bioinformatics, 2018. **34**(18): p. 3169-3177.
13. Hou, H., et al., *Using deep reinforcement learning to speed up collective cell migration.* BMC Bioinformatics, 2019. **20**(18): p. 571.
14. Lillicrap, T.P., et al., *Continuous control with deep reinforcement learning*, in *ICLR*, Y. Bengio and Y. LeCun, Editors. 2016.
15. Watkins, C.J.C.H. and P. Dayan, *Q-learning.* Machine Learning, 1992. **8**(3): p. 279-292.
16. Sutton, R.S. and A.G. Barto, *Reinforcement learning : an introduction*. 1998, Cambridge, Mass. ; London: MIT Press. xviii, 322 p.
17. Silver, D., et al. *Deterministic Policy Gradient Algorithms*. in *ICML*. 2014.
18. Uhlenbeck, G.E. and L.S. Ornstein, *On the Theory of the Brownian Motion.* Physical Review, 1930. **36**(5): p. 823-841.

19. Kingma, D. and J. Ba, *Adam: A Method for Stochastic Optimization.* International Conference on Learning Representations, 2014.
20. Abadi, M., et al., *TensorFlow: a system for large-scale machine learning*, in *Proceedings of the 12th USENIX conference on Operating Systems Design and Implementation*. 2016, USENIX Association: Savannah, GA, USA. p. 265-283.
21. Stukowski, A., *Visualization and analysis of atomistic simulation data with OVITO–the Open Visualization Tool.* Modelling and Simulation in Materials Science and Engineering, 2009. **18**(1): p. 015012.
22. He, K., et al., *Delving Deep into Rectifiers: Surpassing Human-Level Performance on ImageNet Classification*, in *Proceedings of the 2015 IEEE International Conference on Computer Vision (ICCV)*. 2015, IEEE Computer Society. p. 1026-1034.
23. McLennan, R., et al., *Neural crest migration is driven by a few trailblazer cells with a unique molecular signature narrowly confined to the invasive front.* Development, 2015. **142**(11): p. 2014-25.
24. Szabo, A., et al., *Neural crest streaming as an emergent property of tissue interactions during morphogenesis.* PLoS Comput Biol, 2019. **15**(4): p. e1007002.
25. McLennan, R., et al., *Multiscale mechanisms of cell migration during development: theory and experiment.* Development, 2012. **139**(16): p. 2935-44.
25

**Supporting Material**

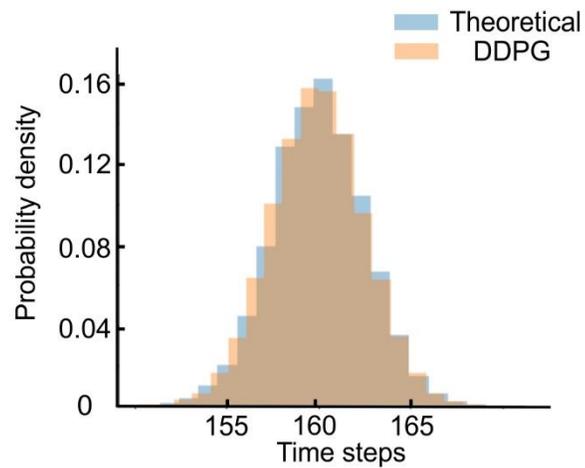

**Fig. S1**. Comparison between the normalized time steps histograms for 10,000 episodes obtained by the DDPG algorithm and the numerical theoretical solution of the diffusion equation.